**Title**

Population pharmacokinetics and dosing regimen optimization of tacrolimus in Chinese lung transplant recipients

**Authors list**

Xiao-Jun Cai [a,c,†], Hui-Zhu Song [a,†], Zheng Jiao [b,*], Hang Yang [d], Min Zhu [b,e], Cheng-Yu Wang [b], Dong Wei [d], Ling-Zhi Shi [d], Bo Wu [d,*], Jing-Yu Chen [d,*]

**Authors' institutional affiliations**

[a] Department of Pharmacy, the Affiliated Wuxi People's Hospital of Nanjing Medical University, Wuxi, 214023, PR China

[b] Department of Pharmacy, Shanghai Chest Hospital, Shanghai Jiao Tong University, Shanghai, 200030, PR China

[c] Department of Pharmacy, Huashan Hospital, Fudan University, Shanghai, 200040, China, PR China

[d] Department of Lung Transplant Center, the Affiliated Wuxi People's Hospital of Nanjing Medical University, Wuxi, 214023, PR China

[e] School of Basic Medicine and Clinical Pharmacy, China Pharmaceutical University, Nanjing, 211198, PR China

**\*Corresponding author**

Professor Zheng Jiao, Ph.D

Department of Pharmacy, Shanghai Chest Hospital, Shanghai Jiao Tong University,

West Huaihai Road 241,

Shanghai, China, 200030.

Tel: +86-21-2220 0000

Email: jiaozhen@online.sh.cn

ORCID: 0000-0001-7999-7162

Dr Bo Wu,

Department of Respiratory, the Affiliated Wuxi People's Hospital of Nanjing Medical University, 299 Qingyang Road, Wuxi, Jiangsu 214002, PR China.

Email: 15852759830@126.com

Professor Jing-Yu Chen, M.D

Department of Lung Transplantation, the Affiliated Wuxi People's Hospital of Nanjing Medical University,

299 Qingyang Road, Wuxi, Jiangsu 214002, PR China.

Email: chenjy@wuxiph.com

ORCID: 0000-0002-2127-1788

† These authors contributed equally to this work.

**Abstract**

We aimed to (i) develop a population pharmacokinetic model of tacrolimus in Chinese lung transplant recipients, and (ii) propose model-based dosing regimens for individualized treatment. We obtained 807 tacrolimus steady-state whole blood concentrations from 52 lung transplant patients and genotyped *CYP3A5\*3*. Population pharmacokinetic analysis was performed using nonlinear mixed-effects modeling. Monte Carlo simulations were employed to determine the initial dosing regimens. Tacrolimus pharmacokinetics was described by a one-compartment model with first-order absorption and elimination processes. In *CYP3A5\*3/\*3* 70-kg patients with 30% hematocrit and voriconazole-free therapy, the mean estimated apparent clearance was 13.1 l h$^{-1}$ with 20.1% inter-subject variability, which was lower than that in Caucasian lung transplant patients (17.5–36.5 l h$^{-1}$). Hematocrit, postoperative days, tacrolimus daily dose, voriconazole concomitant therapy, and CYP3A5*3 genotype were identified as significant covariates for tacrolimus clearance. To achieve target trough concentration (10–15 ng ml$^{-1}$) on the 8$^{th}$ day post-transplant, a higher initial dosage than the current regimen of 0.04 mg kg$^{-1}$ every 12 hours is recommended for *CYP3A5\*1/\*3* patients without voriconazole concomitant therapy. Given the nonlinear kinetics of tacrolimus and large variability, population pharmacokinetic model should be combined with therapeutic drug monitoring to optimize individualized therapy.

# 1. Introduction

Tacrolimus, a potent calcineurin inhibitor, represents the cornerstone prophylaxis treatment for graft rejection in lung transplantation (Miano et al., 2019). Currently, its use in combination with mycophenolate mofetil and corticosteroids can maintain an immunosuppressive regimen in almost 60% of patients, and is preferred over cyclosporine A due to its higher long-term graft and patient survival rates (Christie et al., 2010; European Association for the Study of the Liver. Electronic address, 2016).

Tacrolimus is characterized by a bioavailability of only 25% (range 4%–93%) (Jusko et al., 1995; Vanhove et al., 2016b), owing to its poor aqueous solubility (1–2 μg ml$^{-1}$) and low transmembrane permeability in the intestine mediated by the P-glycoprotein efflux pump (Lee et al., 2016; Vanhove et al., 2016a). After absorption and subsequent entry into circulation, tacrolimus binds to erythrocytes, albumin, and a$_1$-acid glycoprotein (>99%), with a remaining unbound fraction below 1% (Jusko et al., 1995; Yu et al., 2018). It is extensively metabolized by the cytochrome P450 CYP3A (*CYP3A5*) isoenzymes in the liver and predominantly excreted in the feces, with renal excretion accounting for less than 3% of the parent compound (Moller et al., 1999; Staatz and Tett, 2004).

Although the efficacy of tacrolimus is well established, the drug displays a narrow therapeutic window and high intra- and inter-individual pharmacokinetic variability due to multiple factors, such as postoperative time, *CYP3A5*3* genetic polymorphisms, and concomitant medications (Campagne et al., 2019). Signs of rejection and toxicity, including nephrotoxicity, hypertension, and gastrointestinal disorders, induced by tacrolimus are common in transplants, leading to a higher risk of morbidity and mortality (Paradela de la Morena et al., 2010; Wehbe et al., 2013). To optimize the therapeutic effect and minimize tacrolimus-induced toxicity, therapeutic drug monitoring (TDM) is widely used to ensure target whole-blood concentrations, especially during the early post-transplant period. However, the effects of various factors on the pharmacokinetics of tacrolimus have not been considered in TDM.

Population pharmacokinetic (popPK) modeling offers a superior approach compared to other concentration-time measurements derived from TDM, as it can verify and quantify pharmacokinetic variability (Wang et al., 2019). Therefore, a popPK model that contains predictive factors of pharmacokinetic variability may help optimize the initial dose of tacrolimus to improve clinical outcomes. Furthermore, combined with a Bayesian method, it can assist subsequent dosage adjustments in clinical practice (Staatz and Tett, 2011; Storset et al., 2014).

Over the past two decades, several popPK models have described the pharmacokinetics of tacrolimus in solid organ transplant recipients (Staatz and Tett, 2011). Among these models, only three related to lung transplant patients, focusing on stable patients and the pharmacokinetics of whole-blood, total, and unbound plasma tacrolimus in unstable patients early after transplantation with and without cystic fibrosis in Europe (Monchaud et al., 2012; Saint-Marcoux et al., 2005; Sikma et al., 2019). Cystic fibrosis accounts for approximately 16% of lung transplant recipients in Europe (Christie et al., 2010). It also results in a very heterogeneous pharmacokinetic behavior of tacrolimus (Sikma et al., 2019). However, given differences in etiology of cystic fibrosis and genetic background, this disease has not yet become an indication in Chinese lung transplant recipients. Additionally, compared to patients in Europe, Chinese recipients are characterized by an older mean age of 57.28 $\pm$ 18.59 years (range 15–79 years, with those >60 years accounting for more than 35% of cases), more comorbidities, and more risk factors of a poor outcome, including recipient age and preoperative mechanical ventilation (Mao et al., 2013; Wallinder et al., 2019). Therefore, very heterogeneous tacrolimus pharmacokinetic profiles might define Chinese lung transplant recipients, and no dose individualization based on population pharmacokinetic models has been attempted for such patients.

Based on the above considerations, the present study aimed (i) to develop a popPK model that described tacrolimus pharmacokinetics in Chinese lung transplant patients using a retrospectively collected dataset and identify any variability to facilitate dose individualization; and (ii) to perform simulations that optimized dosing regimens during the early stage of lung post-transplantation.

## 2. Methods

### 2.1. Subjects and clinical data collection

Patients at the affiliated Wuxi People's Hospital of Nanjing Medical University, who had their first lung transplantation between September 2015 and June 2018, and were administrated oral tacrolimus to prevent allograft rejection were included in this study. Exclusion criteria comprised serious infections (e.g., septic shock), age below 16 years, acute rejection, dialysis treatment, severe gastrointestinal disorders, or a second lung transplantation.

Whole-blood trough concentrations ($C_0$) of tacrolimus, laboratory tests, and concomitant medications were retrospectively acquired from the patients' medical records. Demographic characteristics of the patients, such as body weight, height, age, and gender, as well as dosing history and sampling time were also collected. The study was approved by the affiliated Wuxi People's Hospital of Nanjing Medical University Ethics Committee for medical research ethics (number: KS202002). All participants gave written informed consent. The study was conducted in accordance with the Declaration of Helsinki (2013).

*2.2. Immunosuppressive therapy*

Patients received an immunosuppressive regimen consisting of tacrolimus, mycophenolate mofetil, and corticosteroids. Oral administration of tacrolimus by immediate-release formulation (Prograf; Astellas Pharma Inc, Dublin, Ireland) was commenced at 0.04 mg kg$^{-1}$ every 12 hours (q12h) under fasting conditions right after transplantation. Subsequent doses were empirically adjusted to achieve $C_0$ of 10–15 ng ml$^{-1}$ in the first 6 months post-operation, 8–12 ng ml$^{-1}$ between 6 and 12 months, and 6–10 ng ml$^{-1}$ thereafter according to routine TDM, clinical evidence of efficacy, and adverse reactions.

Mycophenolate mofetil (CellCept; Roche Pharma Ltd., Shanghai, China) was given orally at 0.5 g q12h for patients with body weight <50 kg and 0.75 g q12h for those ≥50 kg. Intravenous methylprednisolone was administered at 500 mg on the day of transplant and 0.5 mg kg$^{-1}$ day$^{-1}$ during the first three days post-transplant. To maintain the immunosuppressive regimen, prednisolone was given orally at 0.5 mg kg$^{-1}$ day$^{-1}$ on the 4$^{th}$ day post-transplant and was tapered to 0.25 mg kg$^{-1}$ day$^{-1}$ during the first 3 months post-transplant. Thereafter, dosage was tapered to 0.15 mg kg$^{-1}$ day$^{-1}$ until 9 months post-transplant. One year later, corticosteroid-free treatment was administered. Induction therapy was generally used with intravenous basiliximab at 20 mg on the day of transplant and on the 4$^{th}$ day post-transplant.

*2.3. Blood sample collection and analytical method*

Whole-blood samples were drawn before the morning dose and were assessed by a chemiluminescent microparticle immunoassay using the Architect I2000 system (Abbott Laboratories, Abbott Park, IL, USA). Experimental procedures were conducted according to the manufacturer's instructions. Intra- and inter-day coefficients of variation were within 10%, with a calibration range between 2.0 and 30 ng ml$^{-1}$.

*2.4. Genotyping*

Ethylenediaminetetraacetic acid-anticoagulated whole blood (2 ml) was obtained from lung transplant recipients and sent to Sangon Biotechnology Company (Shanghai, China) for DNA extraction, polymerase chain reaction, and subsequent DNA sequencing. Single nucleotide polymorphism of *CYP3A5\*3* (rs776746) was determined by DNA direct sequencing (3730XL; Applied Biosystems, Foster City, CA, USA). Deviation in allele frequencies from the Hardy-Weinberg equilibrium was analyzed using Pearson's chi-square test. Further details are presented in *Appendix 1*.

## 2.5. Population pharmacokinetic modeling

Population pharmacokinetic analysis was performed using a nonlinear mixed-effects modeling software (NONMEM®, version 7.4; ICON Development Solutions, Ellicott City, MD, USA) compiled with gfortran 4.6.0 and interfaced with Perl-speaks--NONMEM (version 4.7.0; uupharmacometrics.github.io/PsN). The NONMEM output was analyzed using R software (version 3.5.1; www.r-project.org). First-order conditional estimation with interaction between inter-patient variability and residual variability was employed for model development.

### 2.5.1. Base model development

Because only trough concentrations were available for analysis, a one-compartment model with first-order absorption and elimination was used to describe the pharmacokinetics of tacrolimus. The model was parameterized in terms of apparent total clearance (CL/F), apparent volume of distribution ($V_d/F$), and absorption rate constant (Ka). As no sampling was performed during the absorption phase, the Ka was fixed at 4.48 $h^{-1}$ based on published data.

Inter-subject variability was estimated for all parameters except Ka, using the exponential model shown in equation (Eq 1):

$$P_i = TV(P) \times exp(\eta_i) \quad \text{(Eq 1)}$$

where $P_i$ is the pharmacokinetic parameter estimation of the *i*th subject and *TV(P)* is the typical value of the population parameter. $\eta_i$ is defined as symmetrically distributed inter-subject variability, with a mean of zero and a variance of $\omega_i^2$.

Residual unexplained variability (RUV) was described using additive (Eq 2), proportional (Eq 3), and combined proportional and additive (Eq 4) models.

$$Y = F + \varepsilon_1 \quad \text{(Eq 2)}$$

$$Y = F + F \ast \varepsilon_1 \quad \text{(Eq 3)}$$

$$Y = F + F \ast \varepsilon_1 + \varepsilon_2 \quad \text{(Eq 4)}$$

Here, Y is the observed concentration, F is the individual predicted concentration, and $\varepsilon_n$ represents symmetrically distributed random variability, with a mean of zero and a variance of $\sigma_n^2$.

*2.5.2 Covariate analysis*

Potential covariates to be investigated included age, gender, height, body weight, hematocrit, albumin, aspartate aminotransferase, total serum bilirubin, creatinine clearance, daily dose of tacrolimus, postoperative days, concomitant medications, and genetic polymorphisms of *CYP3A5*3*. Only comedications >10% in all patients were tested.

Weighted allometric scaling was used to describe CL/F and $V_d$/F, and account for differences in body size and metabolic rate as below (Eq 5):

$$P_i = TV(P) \times (WT_i/70)^{PWR} \qquad (Eq\ 5)$$

where $P_i$ represents the pharmacokinetic parameter of the *i*th individual, $WT_i$ represents the body weight of the *i*th individual, and PWR represents the allometric coefficient for the effect of WT on CL/F and $V_d$/F, which was fixed at 0.75 and 1, respectively (Anderson and Holford, 2008).

Continuous covariates were normalized to the population median values and modeled using linear equations (Eq 6–7) and power (Eq 8) as below:

$$P_i = TV(P) + \theta cov \times (COV_i/COV_{median}) \qquad (Eq\ 6)$$

$$P_i = TV(P) + \theta cov \times (COV_i - COV_{median}) \qquad (Eq\ 7)$$

$$P_i = TV(P) \times (COV_i/COV_{median})^{\theta cov} \qquad (Eq\ 8)$$

where $P_i$ represents the pharmacokinetic parameter of the *i*th individual, *TV(P)* is the typical value for the parameter, $COV_i$ is the covariate value of the *i*th individual, $COV_{median}$ is the population median value of the covariate, and *θcov* is the coefficient term of the covariate effect to be estimated.

The effect of binary covariates such as *CYP3A5*3* polymorphism on parameter P was tested by a scale model (Eq 9):

$$P_i = TV(P) \times (1 + \theta_{CYP} \times CYP) \qquad (Eq\ 9)$$

where *TV(P)* is the typical value for the parameter P in *CYP3A5*3/*3* carriers (*CYP* = 0) and $\theta_{CYP}$ is the fractional change in parameter P in *CYP3A5*1/*1* and *CYP3A5*1/*3* carriers (*CYP* = 1).

Covariates were screened in a stepwise fashion, with forward inclusion and backward elimination. The effect of each variable on parameters was investigated using the likelihood ratio. Statistical significance was set to *P* < 0.05, corresponding to an objective function value (OFV) reduction of at least 3.84 for 1 degree of freedom in forward inclusion, and to *P* < 0.01, corresponding to an OFV reduction of at least 6.63 for 1 degree of freedom for backward elimination. A covariate without a known pharmacological or biological basis, or with <20% effect on a parameter was not retained in the model. Improvement in parameter estimation precision and goodness-of-fit plots, reduction in inter-subject variability and RUV, and stability of the parameter estimates were also considered to select the covariates. Finally, the extent of the shrinkage was evaluated in the final model.

*2.5.3. Model evaluation*

Goodness-of-fit plots, including observed concentrations versus population prediction and individual prediction, the conditional weighted residuals versus population prediction, and time after dose, were used to evaluate the fitness of the final model to the data. Additionally, nonparametric bootstrap, prediction- and variability-corrected visual predictive checks (pvcVPCs), and normalized prediction distribution errors (NPDEs) were employed to evaluate the final model.

For the nonparametric bootstrap procedure, 500 replicate bootstrap data sets were generated from the raw database by random resampling and were fitted with the same model to obtain parameter estimates for each replicate. The medians and $2.5^{th}$–$97.5^{th}$ percentiles of the parameters after bootstrap runs with successful convergence were compared with the final model parameter estimates.

For pvcVPCs, 1000 new data sets were simulated in NONMEM using the final model to produce the expected concentrations. Concentration-time profiles were plotted for the $50^{th}$, $10^{th}$, and $90^{th}$ percentiles of the simulated data and were overlaid with the observed data.

NPDEs were tested with 3000 simulations for each observation in the raw dataset using the final model. NPDE results were summarized statistically and graphically with the NPDE add-on package in R (version 2.0; www.npde.biostat.fr). Their distributions were evaluated to test if the final model fully described the observed data; plots of NPDEs versus observations and NPDEs versus time were also evaluated.

*2.6. Simulation of tacrolimus dosing strategies*

Given how crucial is the target $C_0$ of tacrolimus (10–15 ng ml$^{-1}$) in the first week post-transplant, Monte Carlo simulations were performed to predict the target $C_0$ after 7-day multiple oral doses of different dosing regimens based on the final popPK model. The tacrolimus dose was simulated from 0.5 mg q12h to 6 mg q12h for a 55-kg patient with hematocrit of 30% according to the different *CYP3A5\*3* genotypes and concomitant medication with voriconazole. One thousand simulations were carried out using the initial dataset and steady-state $C_0$ of each simulated subject was calculated.

## 3. Results

### 3.1. Study population

A total of 807 whole-blood trough concentrations of tacrolimus collected from 52 lung transplant patients were available. The reasons for lung transplantation were idiopathic pulmonary arterial hypertension, chronic obstructive pulmonary disease, and idiopathic pulmonary fibrosis. In contrast to previous studies, none of the enrolled patients had a cystic fibrosis diagnosis (Monchaud et al., 2012; Sikma et al., 2019). Forty-eight subjects were sampled at more than three tacrolimus dose levels with a mean of 25 observations per patient. Furthermore, all the recruited patients were co-administered omeprazole and more than half (69.2%) were given regular doses of voriconazole during their therapy. The main demographic characteristics of the enrolled patients are listed in Table 1.

The allele frequencies of *CYP3A5\*3* and *CYP3A5\*1G* genetic polymorphisms are listed in Table 1. There were no deviations from the Hardy-Weinberg equilibrium in these two genotypes ($P > 0.05$). No wild genotype of *CYP3A5\*3* was detected and the frequency of both *CYP3A5\*1/\*3* and *CYP3A5\*3/\*3* was 50% (n = 26).

### 3.2. Population pharmacokinetic modeling

A one-compartment model with first-order absorption and elimination was used as the structural model. The residual error model using a proportional method was selected.

The effect of patient body size on the disposal of tacrolimus was described for CL/F and $V_d$/F by an allometric scaling equation based on weight. The inclusion of body weight as a covariate led to a significant reduction in OFV by 13.499 points. Furthermore, the daily dose of tacrolimus, hematocrit, and postoperative days as continuous covariates, and the co-administration of voriconazole and genetic polymorphisms of *CYP3A5\*3* as discrete covariates, provided a significant decrease of OFV for CL/F. As only trough concentrations were available, the $V_d$/F was not evaluated in the covariate screening. In the backward elimination step, all the covariates were retained in the final model. The results of stepwise forward addition and backward elimination are shown in Table S1.

The final model was as follows:

$$\text{CL/F} = 13.1 \times (\text{WT}/70)^{0.75} \times (\text{HCT}/30)^{-0.868} \times (\text{DD}/3)^{0.616} \times (\text{POD}/30)^{0.0807}$$

$$\times 1.3 \text{ (if } CYP3A5\text{*}1 \text{ carriers)}$$

$$\times 0.638 \text{ (if with voriconazole comedication)} \quad \text{(Eq 11)}$$

$$V_d/F = 636 \times (\text{WT}/70) \quad \text{(Eq 12)}$$

where WT is the body weight, HCT is the hematocrit, DD is the daily dose of tacrolimus, POD is the postoperative days, and voriconazole is co-therapy with voriconazole.

The population parameter estimates of CL/F and $V_d$/F in the final model were 13.1 l h$^{-1}$ and 636 l, respectively. Compared to the base model, all the identified covariates in the final model explained the 38.4% inter-subject variability and 13.9% RUV in CL/F. Among covariates, *CYP3A5\*1/\*3* genetic polymorphism presented an obvious effect on CL/F; specifically, CL/F was 30.0% faster in patients with *CYP3A5\*1/\*3* than in those with *CYP3A5\*3/\*3*. Additionally, voriconazole concomitant therapy contributed to a 36.2% lower CL/F compared to no treatment. The final model was characterized by clearance showing 20.1% inter-subject variability and 32.9% RUV. Shrinkage was estimated at 14.2% for CL/F and 7.7% for $V_d$/F. All parameter precision represented by a standard error was acceptable. *Table 2* shows the population pharmacokinetic parameter estimates and precision of the final model with covariates.

### 3.3. Model evaluation

The goodness-of-fit plots for the base and final model are shown in *Figure 1*. No significant structural bias and obvious systematic deviations were found in the final model, and data fitting was greatly improved compared to the base model.

Bootstrap analysis was successful in 93.4% of the 500 runs. The median values of the pharmacokinetic parameter and 2.5%–97.5% estimates obtained from bootstrap were close to those obtained with the original dataset with <5% bias, confirming the stability and robustness of the final model (*Table 2*).

The pvcVPCs of the final model are presented in *Figure 2*. The 10$^{th}$, 50$^{th}$, and 90$^{th}$ percentiles of the observations were within the 95% confidence intervals of the corresponding prediction percentiles for the final model, revealing an acceptable agreement between the simulated and observed concentrations at all sampling time points.

The NPDE results are depicted in *Figure 3*. The quantile-quantile plots and histogram confirmed the normality of the NPDE and the good predictability of the final model. The assumption of a normal [0,1] distribution for the differences between predictions and observations was acceptable with adjusted *P* values of 0.0543 for the global test.

### 3.4. Dosing strategies

The results of Monte Carlo simulations are shown in *Figure 4*. Receiving the current starting dose of 0.04 mg

kg$^{-1}$ q12h, 55-kg patients with *CYP3A5\*3/\*3* or *CYP3A5\*1/\*3* carriers, a hematocrit of 30%, and comedication with voriconazole, could reach a target steady-state $C_0$ of 10–15 ng ml$^{-1}$ on the 8$^{th}$ day post-transplant. In contrast, for *CYP3A5\*1/\*3* carriers without voriconazole comedication, the current dosage regimen might be insufficient. Moreover, to achieve the target concentration, patients co-administered with voriconazole received a lower maintenance dose than *CYP3A5* non-expressing patients receiving only tacrolimus (1.0–2.0 mg q12h vs. ≥2.0 mg q12h). Under the designed dosing regimens, over 50% of patients could not achieve the target tacrolimus $C_0$ range.

## 4. Discussion

To the best of our knowledge, this is the first popPK model for tacrolimus in Chinese lung transplant patients. Body weight, hematocrit, postoperative days, daily dose of tacrolimus, *CYP3A5\*3* polymorphism, and concomitant medication with voriconazole were identified as covariates of CL/F. Our study shows that for a typical 55-kg patient with *CYP3A5\*3/\*3*, with a daily tacrolimus dose of 3 mg, a median of 27 postoperative days, and hematocrit of 30%, the CL/F was estimated to be 13.1 l h$^{-1}$, which is much lower than the published value (17.5–36.5 l h$^{-1}$) for a non-Asian population (Monchaud et al., 2012; Sikma et al., 2019). This discrepancy could be explained by the median age of the subjects (54 years, range 16–78 years) being typically higher than that of previous studies (41–43 years) (Monchaud et al., 2012; Sikma et al., 2019). Aging is associated with a gradual decrease in clearance due to reduced total body water and lean body mass, and a relative increase in body fat (Fulop et al., 1985). These changes result in a larger volume available for hydrophobic tacrolimus. Additionally, changes in intestinal motility and gastric pH, as well as decreased liver blood flow in older patients contribute to a lower clearance of tacrolimus, especially in the first 6 months post-transplantation (Jacobson et al., 2012).

Differences in immunosuppressant regimens, including steroid tapering regimen, concomitant medications, and food constituents across medical centers, could lead to variations in the drug's disposal by influencing the action of CYP3A and bioavailability (Fu et al., 2019; Vanhove et al., 2016a). Also, as the analytical method may result in bias of parameters estimation (Akbas et al., 2005), the method for assaying tacrolimus $C_0$ may be inconsistent with that used in previous studies (Monchaud et al., 2012; Sikma et al., 2019) and give rise to the lower estimated CL/F (Brunet et al., 2019). Furthermore, ethnic variations between Chinese and Caucasian patients (Staatz et al., 2010; Tang et al., 2016) might be another factor. Polymorphic expression and ethnic-specific distribution or activity of *CYP3A5 and ABCB1* alleles have been suggested to play an important role in the ethnic variations observed in tacrolimus pharmacokinetics (Campagne et al., 2019; Dirks et al., 2004). In this as well as in previous studies on lung transplantation, only the impact of *CYP3A5*3* genotypic variants has been explored; *CYP3A5*6* and *7* variants and *ABCB1* genotypes with inter-ethnic comparisons have not been included.

Interestingly, the incorporation of daily drug dose as a covariate significantly improved data fitting by decreasing the OFV by 298.208 points, indicating nonlinear clearance of the drug in lung transplant patients. The CL/F of tacrolimus increased nonlinearly with higher daily drug doses, which is in agreement with previous reports in kidney and liver transplant recipients (Ahn et al., 2005; Zhao et al., 2016). Such nonlinear kinetics can be explained by poor water solubility and low transmembrane permeability of tacrolimus in the intestine. These properties cause a dissolution rate-limited absorption in the gastrointestinal tract, as well as variable and low oral bioavailability (Lee et al., 2016). Further research with a larger sample size and intensive sampling will provide a more specific picture of the nonlinear behavior of tacrolimus kinetics in lung transplantations.

Weight was identified as a significant covariate on tacrolimus pharmacokinetics in lung transplant patients. A positive relationship was found between body weight and CL/F and $V_d$/F, indicating that a higher weight-normalized dose of tacrolimus was required at increasing body weight.

A negative relationship was found here between tacrolimus CL/F and the hematocrit after lung transplantation. This may be explained by the strong binding of tacrolimus to erythrocytes, facilitated by the drug's lipophilicity and the presence of immunophilins in red blood cells (Vanhove et al., 2016a). A lower hematocrit commonly leads to a transiently increased unbound fraction of tacrolimus, resulting in a higher clearance and subsequent higher tacrolimus daily dose requirement as more tacrolimus is metabolized and eliminated from the body.

Postoperative days has been identified as a major surrogate for many time-dependent variables (Pou et al., 1998). Here, a gradual improvement in metabolic function with postoperative days increased the CL/F after lung transplantation. This may be attributed to the progressive recovery of *CYP3A* enzyme activity in the intestine and liver and the tapering of corticosteroid doses with increasing time after transplantation, particularly during the initial postoperative period. Other factors and possible mechanisms related to the effect of postoperative days on tacrolimus pharmacokinetics remain to be determined.

Various studies have shown that *CYP3A5*3* polymorphism is associated with the pharmacokinetics of tacrolimus (Wang, 2009; Zheng et al., 2004). Our observation further confirmed these findings and demonstrated that CL/F of tacrolimus for lung transplant patients was 1.3-fold higher in *CYP3A5*1* carriers than in non-carriers (*CYP3A5*3/*3*), which is similar to the published value of 1.4-fold (Monchaud et al., 2012). Introduction of this covariate accounted for 10% of inter-subject variability in tacrolimus clearance, which is close to the value of 6% reported in the literature (Jacobson et al., 2011). Therefore, given the wide availability of TDM, genetic testing prior to initiating tacrolimus treatment is important for dosage optimization.

Besides genetic factors, concomitant medications usually lead to drug-drug interactions involving tacrolimus. These contribute to the observed variability of pharmacokinetics in lung transplantations as they affect CYP3A- and P-glycoprotein-mediated metabolism (Christians et al., 2002). Voriconazole, an inhibitor of CYP3A activity in both the intestine and the liver, is frequently administered to lung transplant recipients (Huang et al., 2020). Here, its administration led to a 36.2% slower clearance relative to voriconazole-free patients. Its inhibition of intestinal and hepatic CYP3A enzymes reduced tacrolimus clearance, resulting in a higher concentration of tacrolimus, which is consistent with findings in patients with primary nephrotic syndrome (Huang et al., 2020). Thus, TDM should be conducted before the administration of voriconazole, and the dosage regimen should be adjusted according to the target tacrolimus concentration.

Wuzhi capsule and calcium channel blockers are extensively used as a tacrolimus-sparing agent (Li et al., 2011). They were not included as covariates in the final model due to the small patient size. Corticosteroids are also regarded as important inducers of CYP3A and P-glycoprotein, and are likely to affect the disposal of tacrolimus. However, they were also not included in the final model, probably due to the low dose of co-administered steroids ($18.67 \pm 25.95$ mg day$^{-1}$ in the form of prednisone).

For the population in this study, most concentrations were taken at steady-state trough time-points and in the clinic, where the dose of tacrolimus is normally adjusted according to the monitored $C_0$ (10–15 ng ml$^{-1}$). Therefore, the current study used the developed final model to predict the $C_0$ of tacrolimus and test the possible clinical impact of *CYP3A5* genotype and concomitant medication with voriconazole. Simulation results demonstrated a high variability in tacrolimus pharmacokinetics, whose origin remains unknown. Moreover, they indicated that the present initial dose of 0.04 mg kg$^{-1}$ q12h was insufficient for a 55-kg patient with hematocrit of 30%, *CYP3A5*1/*3,* and receiving voriconazole-free treatment to reach the target concentration range.

There are some limitations in this study. The first and most obvious one is the relatively small sample size. A second limitation is the incorporation of trough concentrations only, which were employed to reliably assess CL/F and its covariates. A third limitation is the lack of an evaluation of the relationship between efficacy or toxicity and tacrolimus trough concentrations. Linking the pharmacokinetics to clinical outcomes with the aim of targeting a more appropriate therapeutic range remains a critical but challenging aspect of tacrolimus therapy. Despite all the above limitations, the study can provide the theoretical basis for personalized tacrolimus treatment and serve as a valuable reference for lung transplantations.

## 5. Conclusion

In this study, a popPK model to characterize the pharmacokinetics of tacrolimus was developed for the first time in Chinese lung transplant patients. Significant covariates, including hematocrit, postoperative days, daily dose of tacrolimus, concomitant medication with voriconazole, and *CYP3A5*3* genotypes, were identified in the final model. This study highlights a lower tacrolimus clearance in Chinese lung transplant patients than in Caucasians. The discrepancy might be attributed to ethnic variation and different age composition of the samples. Another important discovery is the identification of an underlying nonlinear kinetics of tacrolimus. *CYP3A5*1/*3* 55-kg patients with 30% hematocrit and voriconazole-free therapy require a higher initial dose than the current regimen (0.04 mg kg$^{-1}$ q 12 h) to achieve the target concentration. The unknown high variability in tacrolimus pharmacokinetics revealed by the simulation results suggests that popPK model should be combined with traditional TDM to optimize dose regimens to reach target trough tacrolimus levels more effectively.

**Acknowledgements**

This work was supported by the "Weak Discipline Construction Project" (2016ZB0301-01) of Shanghai Municipal Commission of Health and Family Planning. We would also like to thank Editage (www.editage.cn) for English language editing.

**Declaration of Competing Interest**
None

**Author contributions**
Xiaojun Cai, Zheng Jiao, and Jingyu Chen designed the study and planned the work that led to the manuscript. Xiaojun Cai, Huizhu Song, Bo Wu, and Lingzhi Shi contributed to acquisition of data and interpreted data. Xiaojun Cai, Hang Yang, Lingzhi Shi collected the blood samples and performed pharmacological and genotyping analyses. Xiaojun Cai, Chengyu Wang, and Min Zhu analysed the data and provided statistical expertise. Xiaojun Cai drafted the manuscript and Zheng Jiao revised the manuscript. All authors approved the final version to be published.

**Figure legends**

**Figure 1** Goodness-of-fit plots for the base model and the final model. A. Observed versus population predicted concentration, B. observed versus individual predicted concentration, C. conditional weighted residuals (CWRES) versus population predictions, and D. CWRES versus time after dose. The solid lines in A and B are identity lines, and the solid lines in C and D are zero lines.

**Figure 2** Prediction- and variability-corrected visual predictive check (pvcVPC) plot of the final pharmacokinetic model. The red solid line represents the prediction- and variability-corrected median observed concentration, and the semitransparent red shaded area represents the simulation-based 95% confidence intervals (CIs) for the median. The red dashed lines represent the corrected observed 10th and 90th percentiles, and the semitransparent blue shaded areas represent the simulation-based 95% CIs for the corresponding predicted percentiles from the final model. The blue dots represent the prediction- and variability-corrected observations.

**Figure 3** Normalized prediction distribution error (NPDE) plot of the final model. A. Quantile-quantile plot of the distribution of the NPDE against the theoretical distribution (semitransparent blue fields), B. Histogram of the distribution of the NPDE against the theoretical distribution (semitransparent blue fields), C. NPDE vs. postoperative time (days), D. NPDE vs. predicted concentrations. In plot C and D, the red solid lines represent the median NPDE of the observations, and semitransparent red fields represent the simulation-based 95% confidence intervals (CIs) for the median. Blue solid lines represent the NPDE of the observed $5^{th}$ and $95^{th}$ percentiles, and semitransparent blue fields represent the simulation-based 95% CIs for the corresponding predicted percentiles from the final model. The blue dots represent the NPDE of the observations.

**Figure 4** Boxplots of the distributions of simulated tacrolimus trough concentrations for *CYP3A5*1/*3* and *3/*3* with co-administered voriconazole on 0.5, 1, 1.5, 2, 2.5, and 3 mg q12h, and for non-combined voriconazole group on 1, 2, 3, 4, 5, and 6 mg q12h regimens in lung transplant patients. The bold horizontal bars in the middle show the median values, whereas the outer boundaries of the boxes represent the ranges of the $25^{th}$ and $75^{th}$ percentiles (interquartile ranges). The whiskers indicate the maximum and the minimum values of trough concentrations.

**Tables**

**Table 1 Characteristics of lung transplant recipients used for model development**

| Characteristics | Number or Mean ± SD | Median (Range) |
| --- | --- | --- |
| No. of patients (Male/Female) | 52 (37/15) | / |
| No. of tacrolimus samples | 807 | / |
| Age (years) | 51.66±15.04 | 54 (16-78) |
| Height (m) | 1.67±0.08 | 1.69 (1.43-1.83) |
| Total body weight, WT (kg) | 53.77±10.9 | 55 (32-75) |
| Predicted fat free mass, FFM (kg) [a] | 43.16±9.33 | 46.17 (20.87-59.9) |
| Hemoglobin, HB (g $l^{-1}$) | 99.16±16.52 | 98 (21-149) |
| Haematocrit, HCT (%) | 29.71±4.58 | 29.3 (18-41.7) |
| Total serum protein, TP (g $l^{-1}$) | 61.2±7.23 | 60.5 (43.5-92.2) |
| Serum albumin, ALB (g $l^{-1}$) | 38.47±6.35 | 38 (24.6-59.5) |
| Alanine transaminase, ALT (U $l^{-1}$) | 20.01±21.88 | 14 (2-289) |
| Aspartate aminotransferase, AST (U $l^{-1}$) | 21.75±25.59 | 16 (5-348) |
| Alkaline phosphatase, ALP (U $l^{-1}$) | 71.3±41.87 | 65 (20-757) |
| γ-Glutamyl transpeptidase, γ-GGT (U $l^{-1}$) | 36.5±49.01 | 20 (4-678) |
| Total serum bilirubin, TBIL (μmol $l^{-1}$) | 18.84±14.67 | 14.6 (3.8-157.5) |
| Blood uric nitrogen, BUN (mmol $l^{-1}$) | 9.01±5.45 | 7.5 (0.5-42.6) |
| Serum creatinine, SCR (μmol $l^{-1}$) | 73.11±42.62 | 63.9 (19.9-561.3) |
| Creatinine clearance, CCR (ml $min^{-1}$) [a] | 89.16±38.56 | 83.64 (11.06-289.29) |
| Methylprednisolone dose (mg $day^{-1}$) | 18.67±25.95 | 16(6-524) |

*Continues*

**Table 1 continued**

| Characteristics | Number or Mean ±SD | Median (Range) |
| --- | --- | --- |
| Sampling time (Postoperative days), POD (day) | 35.79±30.62 | 27 (2-162) |
| Tacrolimus daily dose (mg day$^{-1}$) | 3.8±2.58 | 3 (0.125-13) |
| Tacrolimus trough concentration (ng ml$^{-1}$) | 10.82±4.59 | 10.3 (1.2-29.7) |
| Concomitant medication [b] | | |
|     Wuzhi capsule | 11 | / |
|     Calcium channel blocker | 13 | / |
|     Voriconazole | 285 | / |
|     Methylprednisolone | 807 | / |
|     Mycophenolate mofetil | 191 | / |
| Genotype, n (%) [c] | | |
| CYP3A5*3 (A6986G, rs776746) | | |
|     GA (*1/*3) | 26 (50.0%) | / |
|     GG (*3/*3) | 26 (50.0%) | / |

SD, standard deviation

[a] Calculated from serum creatinine using the Cockcroft-Gault formula: CCR= [140 – age (years)] × weight (kg) / [0.818 × SCR (μmol l$^{-1}$)] × (0.85, if female)

[b] Data are expressed as number of samples

[c] All allele frequencies were in agreement with those predicted by Hardy-Weinberg equation ($P > 0.05$).

**Table 2** Population pharmacokinetic parameter estimates of the base model, final model, and bootstrap

| Parameter | Base model | | Final model | | Bootstrap of final model | | Bias (%) |
|---|---|---|---|---|---|---|---|
| | Estimate | RSE (%) | Estimate | RSE (%) | Median | 2.5%-97.5% | |
| $k_a$ (h$^{-1}$) | 4.48 | / | 4.48 | / | 4.48 | / | / |
| CL*(Body weight/70)$^{0.75}$/F (l h$^{-1}$) | 9.93 | 8.4 | 13.1 | 4.8 | 13.1 | 11.8-14.5 | 0.00 |
| *CYP3A5*1*3* | / | / | 0.3 | 28.2 | 0.310 | 0.158 - 0.499 | 3.33 |
| Haematocrit | / | / | -0.868 | 18.5 | -0.871 | -1.202 - (-0.504) | 0.34 |
| Postoperative days | / | / | 0.0807 | 29.0 | 0.084 | 0.036 - 0.129 | 4.09 |
| Daily dose of tacrolimus | / | / | 0.616 | 8.4 | 0.613 | 0.518 - 0.706 | -0.49 |
| Co-therapy with Voriconazole | / | / | -0.362 | 11.8 | -0.366 | -0.451 - (-0.248) | 1.10 |
| $V_d$* (Body weight/70)/F (l) | 1160 | 13.6 | 636 | 15.3 | 642 | 435 - 960 | 0.94 |
| Between-subject variability | | | | | | | |
| CL/F (%) | 58.5 | 9.9 | 20.1 | 14.9 | 19.3 | 13.7 - 24.9 | -3.98 |
| Shirinkage (%) | 3.4 | / | 14.2 | / | / | / | / |
| $V_d$/F (%) | 93.8 | 11.1 | 98.2 | 9.9 | 96.4 | 68.5 - 113.7 | -1.83 |
| Shirinkage (%) | 8.5 | / | 7.7 | / | / | / | / |
| Residual variability | | | | | | | |
| Proportional error (%) | 46.8 | 8.9 | 32.9 | 8.1 | 32.6 | 29.8 - 35.4 | -0.91 |
| Shirinkage(%) | 4.0 | / | 4.1 | / | / | / | / |

RSE, relative standard error; Bias% = (Bootstrap – NONMEM) / NONMEM ×100%

Figure 1

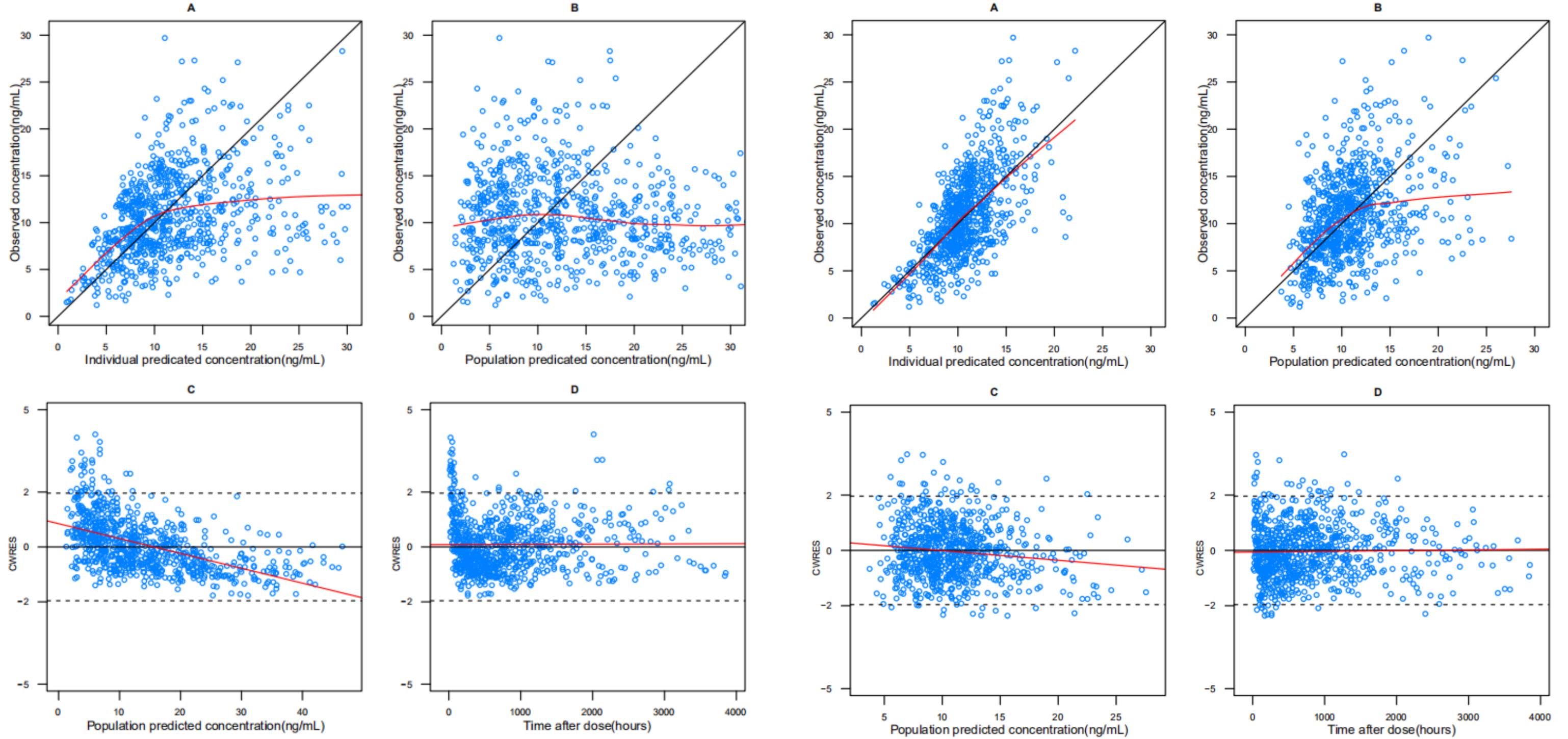

**Figure 2**

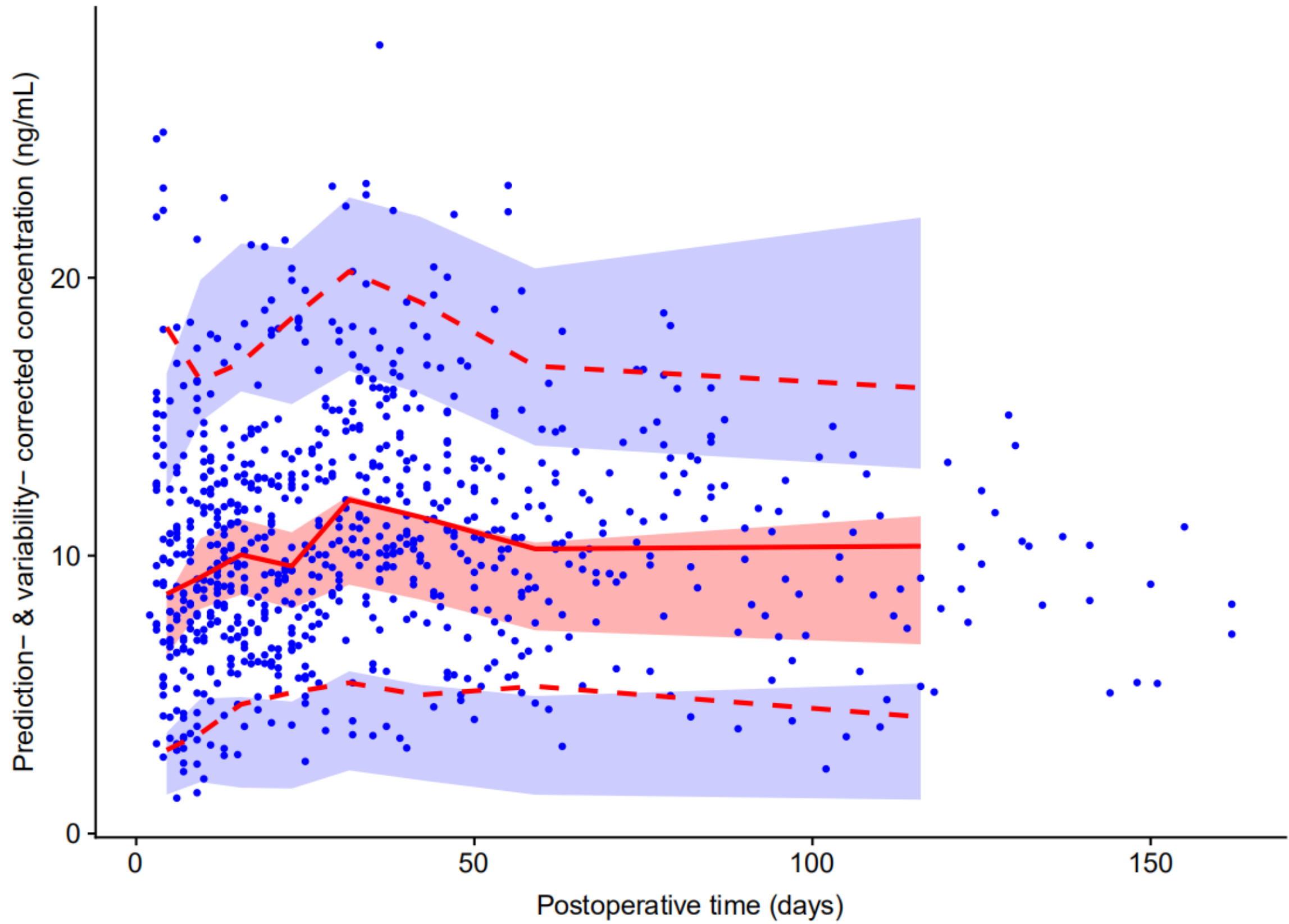

**Figure 3**

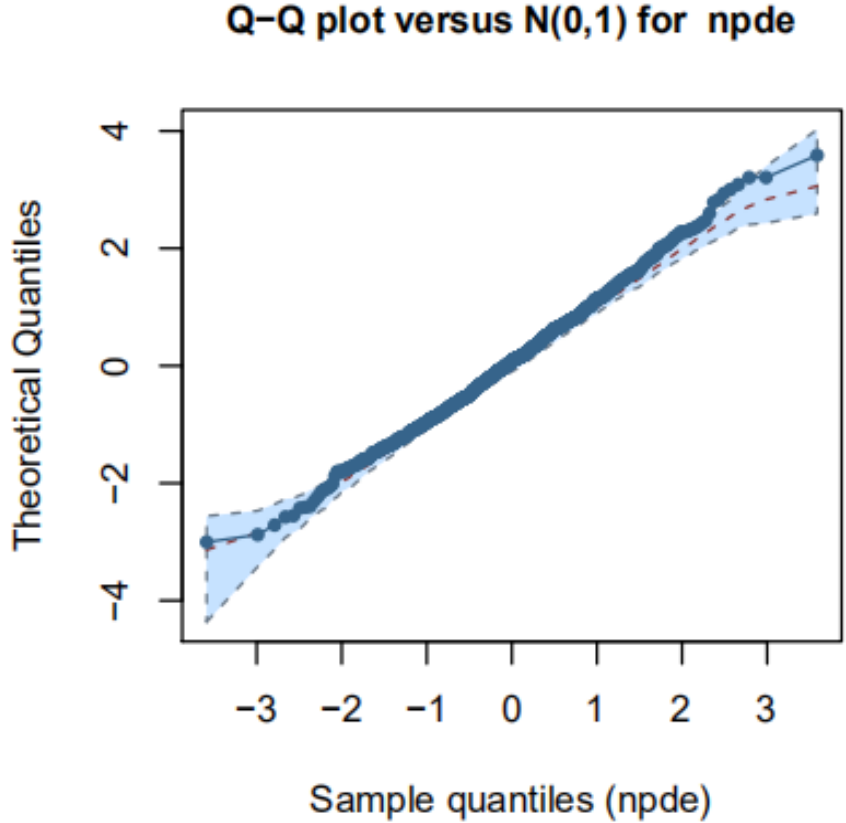
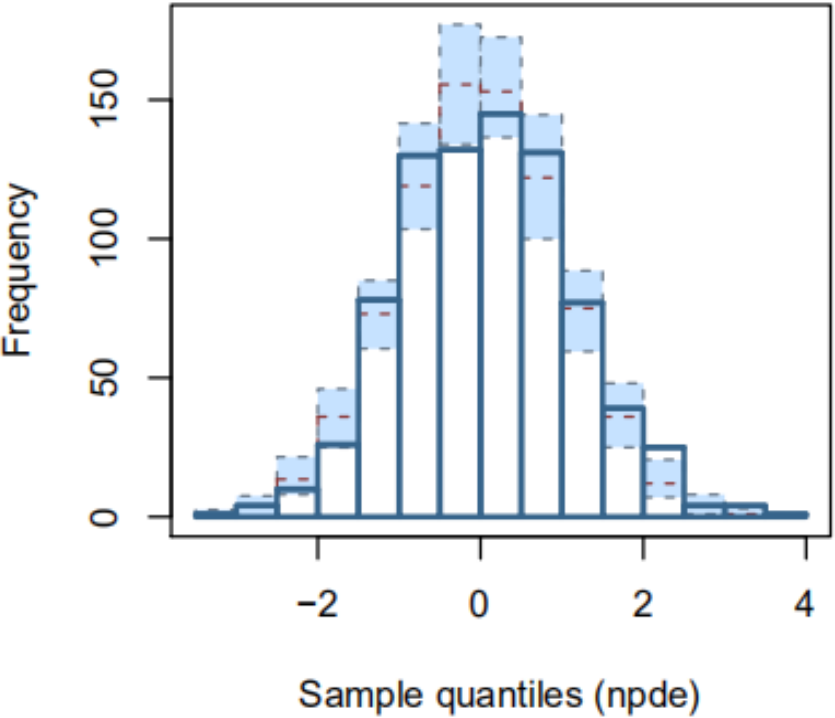
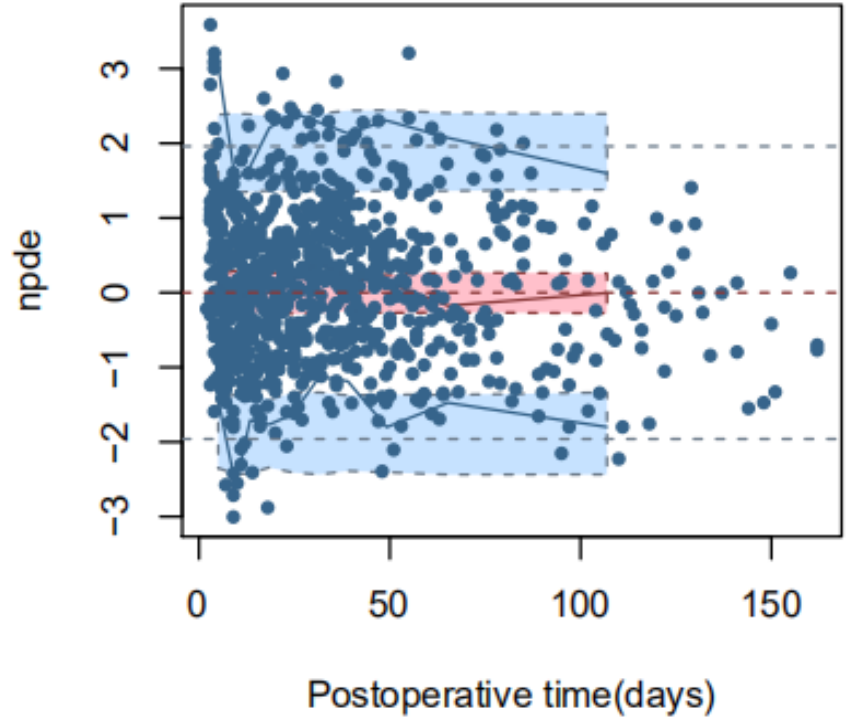
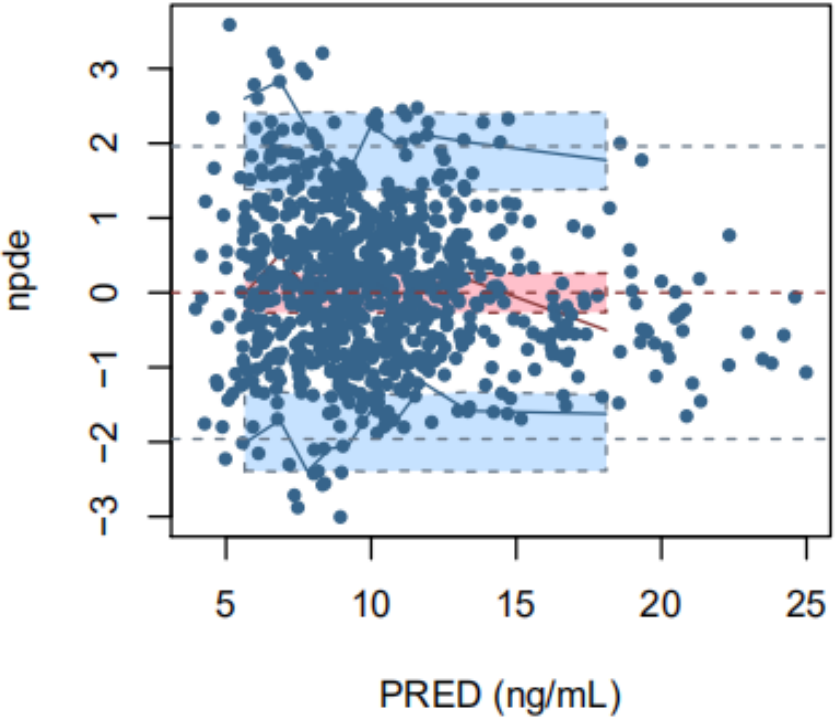

**Figure 4**

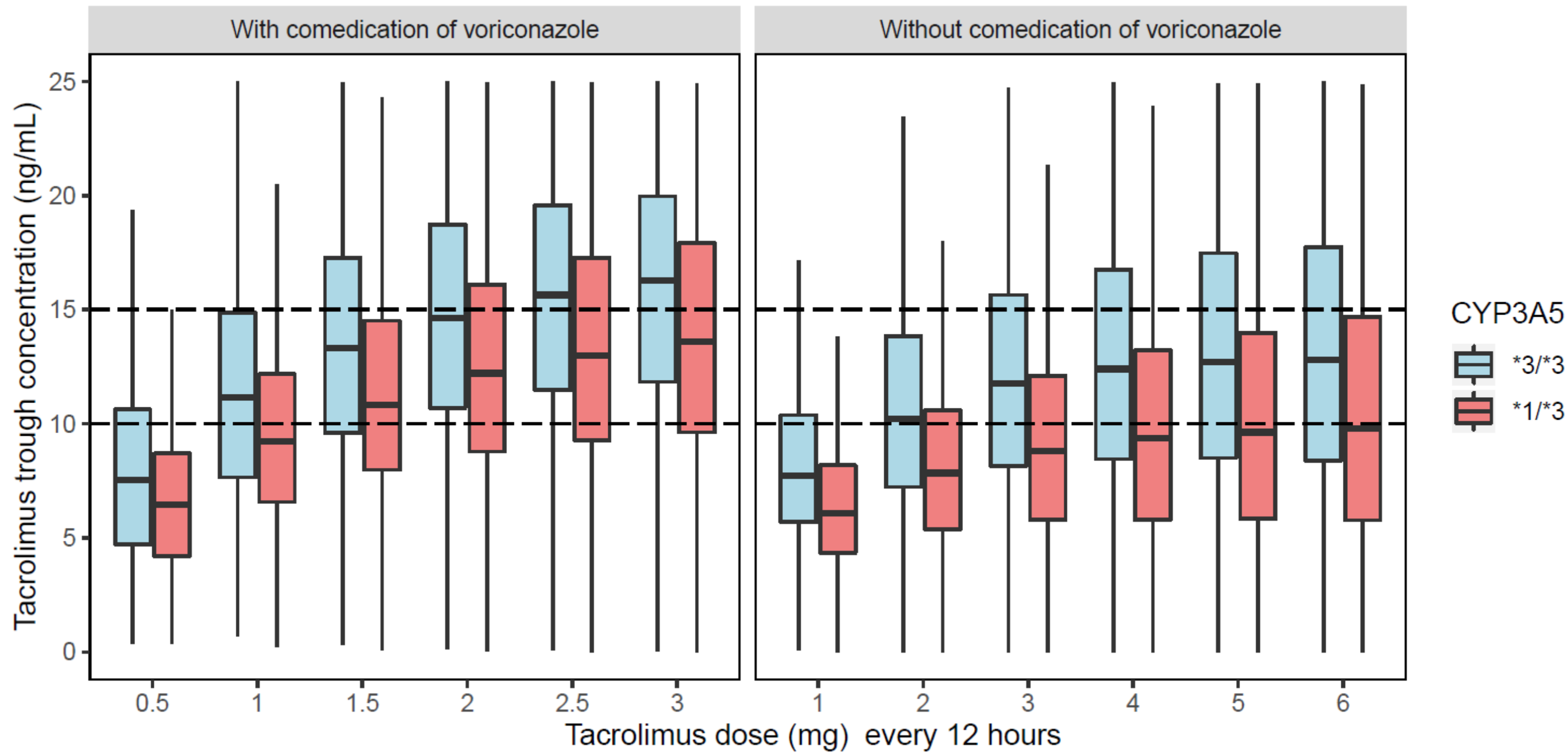

*Supporting information*

Additional Supporting Information may be found in the online version of this article at the publisher's web-site.

*Appendix S1*

Genotyping of *CYP3A5*3* single-nucleotide polymorphisms

*Table S1*

Results of forward inclusion and backward elimination process

*Appendix S1. Genotyping of CYP3A5*3 single-nucleotide polymorphisms.*

DNA was extracted from the whole blood of both liver transplant receipients and their corresponding donors using the TIANamp Blood DNA Kit (Tiangen Biotech Co. Ltd, Beijing, China). The polymerase chain reaction (PCR) was applied to amplify the variant alleles using ABI Veriti 96-Well PCR (Applied Biosystems, Foster City, CA, USA). The volume of amplification reaction was 20 μl, containing 10 μl 2×Taq Master Mix, 7 μl of distilled water, 1 μl of each primer (10 pmol μl$^{-1}$), and 1 μl DNA template (20 ng μl$^{-1}$). The sequences of forward (F) and reverse (R) primers were listed below.

**The sequences of forward (F) and reverse (R) primers for the CYP3A5 genotyping SNPs**

| Gene | Primer | $T_m$ |
| --- | --- | --- |
| *CYP3A5*3* | F: 5' CATTTAGTCCTTGTGAGCACTTGAT 3' | 59.9 ℃ |
| (rs776746) | R: 5' TAGCACTGTTCTGATCACGTCG 3' | 58.8 ℃ |

The PCR conditions were as follows: a denaturation at 95 °C for 3 minutes, then 35 cycles of denaturation at 94 °C for 30 seconds, annealing at 55 °C for 25 seconds, and elongation at 72 °C for 30 seconds, followed by a final extension at 72 °C for 5 minutes. The amplified DNA was purified and genotypes were determined by direct sequencing using ABI PRISM 3730XL Sequence Detection System (Applied Biosystems, Foster City, CA, USA). Tm, melting temperature.

*Supplementary table*

Table S1   Forward inclusion and backward elimination process

| Model number | Model description | OFV | ΔOFV | Significant |
|---|---|---|---|---|
| 1 | **Base model** | 3730.767 | | |
| | **Forward inclusion** | | | |
| 2 | Add WT on CL and V in model 1 | 3717.092 | -13.675 | YES |
| 3 | ADD HCT on CL in model 2 | 3613.327 | -103.765 | YES |
| 4 | ADD CYP3A5 on CL in model 3 | 3594.822 | -18.505 | YES |
| 5 | ADD co-therapy with voriconazole on CL in model 4 | 3334.359 | -260.463 | YES |
| 6 | ADD POD on CL in model 5 | 3285.589 | -48.77 | YES |
| 7 | ADD DD on CL in model 6 | 2987.381 | -298.208 | YES |
| 8 | ADD ALB on CL in model 7 | 2986.457 | -0.924 | NO |
| | **Backward elimination** | | | |
| 9 | Remove HCT on CL from model 7 | 3055.428 | 68.047 | YES |
| 10 | Remove CYP3A5 on CL from model 7 | 2994.707 | 7.328 | YES |
| 11 | Remove co-therapy with voriconazole on CL from model 7 | 3079.030 | 91.649 | YES |
| 12 | Remove POD on CL from model 7 | 3016.719 | 29.338 | YES |
| 13 | Remove DD on CL from model 7 | 3285.589 | 298.208 | YES |